\documentclass[fleqn,twoside]{article} \usepackage{espcrc2}

\usepackage{graphicx,color}
\usepackage[figuresright]{rotating}

\newcommand{\AmS}{{\protect\the\textfont2
A\kern-.1667em\lower.5ex\hbox{M}\kern-.125emS}}

\title{The heavy quark's self energy from moving NRQCD on the lattice}

\author{A. Dougall\address{Department of Physics and Astronomy,
    University of Glasgow, Glasgow G12 8QQ, UK},
        C. T. H. Davies\addressmark,
        K. M. Foley\address{Newman Laboratory 
	  for Elementary-Particle Physics,
        Cornell University, Ithaca NY 14853, USA},
        G. P. Lepage\addressmark,
HPQCD and UKQCD collaborations}
       
\begin{document}

\begin{abstract}
We present a calculation of the heavy quark's self energy in moving
NRQCD to one-loop in perturbation theory. Results for the energy shift
and external momentum renormalisation are discussed and compared with
non-perturbative results. We show that the momentum renormalisation is
small, which is the result of a remnant of re-parameterisation invariance
on the lattice.

\vspace{1pc}
\end{abstract}

\maketitle

\section{INTRODUCTION}

The study of semi-leptonic $B$ decays is currently of great interest
as it provides us with a test of the Standard Model. A key process
is that of $B\rightarrow \pi$ semi-leptonic decay, however, because
the $B$ meson is much heavier than the final states, the recoil
momenta of the $\pi$ meson can be large. This leads to both problems
of small signal to noise ratios and significant discretisation errors
if $ap_\pi$ is large. Computations involving the simulation of heavy
quarks at low momentum can be dealt with using the NRQCD formalism
\cite{nrqcd}, where the dynamics of interest are isolated by
re-expressing the Lagrangian as an expansion in $v/c$. Although this
framework for dealing with heavy quarks reduces the impact of
discretisation errors, it is not designed to deal with large
momenta. In this case, a suitable formalism to use is moving NRQCD
(mNRQCD) \cite{mnrqcd1,mnrqcd2,mnrqcd3} which allows the momentum to be
shared more evenly between the initial and final states. mNRQCD is
developed in two stages, the first of which involves a choice of
lattice frame such that the $B$ meson is moving, thus reducing the
magnitude of the pion's recoil momentum. Secondly, an effective
field  theory must be constructed for the heavy quark with non-zero
velocity.  In mNRQCD, the heavy quark's momentum is written as
\begin{eqnarray*}
  P_Q^\mu = M_Q u^\mu + k^\mu
\end{eqnarray*}

\noindent
where $u^\mu = \gamma (1,{\bf v})$ is the 4-velocity of the $b$ quark which
is treated exactly, and $k^\mu$ is the small internal momentum (of the
heavy quark) inside the $B$ meson which is discretised on the lattice.

As with the non-moving case, mNRQCD is constructed from a set of
non-renormalisable interactions, each of which is accompanied by a
coupling coefficient which can be calculated perturbatively. The heavy
quark's self energy in NRQCD has been calculated to $O(\alpha_s)$
\cite{morn}. Here we present a calculation of the one-loop
renormalisation parameters that appear in the mNRQCD Lagrangian. In
addition to the overall energy shift, mass and wavefunction
renormalisation, we must also now calculate the renormalisation of the
external momentum. This renormalisation is close to one because of
the remnants of re-parameterisation invariance on the lattice.

\section{COMPUTING THE SELF ENERGY}

In this section, we present the Lagrangian used in this calculation
and define the renormalisation parameters.

\subsection{Perturbation theory}

The Lagrangian is defined as 
\begin{eqnarray}
  a\mathcal{L}_{\mathrm{NRQCD}} =
  \psi^{\dagger}(x)\psi(x)-\,
  \psi^{\dagger}(x+a_t)\, \mathrm{H} \, \psi(x)
\label{l}
\end{eqnarray}

\noindent
where
\begin{eqnarray*}
  \mathrm{H} = \frac{1}{u_0} \left(1-\frac{aH_0}{2n}\right)^{n}
  U^{\dagger}_t(x)
  \left(1-\frac{aH_0}{2n}\right)^{n}
\end{eqnarray*}

\noindent
and
\begin{eqnarray*}
  H_0 = -\frac{1}{2\gamma M } \left(\Delta^{(2)} -
  \left[{\bf v}.{\bf \nabla}  \right]^2 \right) -
  i{\bf v}.{\bf \nabla}
\end{eqnarray*}

\noindent
In Eqn.~\ref{l}, $n$ is a positive parameter that stabilises the high
momentum modes in the heavy quark propagator. The mean-field parameter
$u_0$ is used to improve the match between lattice and continuum
theories. The gluon fields are discretised using the Wilson
plaquette gauge action. This Lagrangian includes all spin-independent 
corrections up to $O(1/M,v^2)$.

The Feynman rules are calculated by expanding the $U$ fields in terms
of the coupling $g$. Following a Fourier transform to momentum space,
the quark (gluon) propagator is defined as the inverse of the term
that is bilinear in the quark (gluon) fields. Vertices of different
order are selected according to their order in $g$. The rules are
defined in the Feynman gauge and the infra-red divergence is regulated
by introducing a small gluon mass. Note that in the limit $v
\rightarrow 0$, we recover the Feynman rules for the 
corresponding NRQCD action.

\subsection{Outline of the calculation}
\begin{center}
\begin{figure}
\vspace{.6cm}
\hspace{.6cm}
  \includegraphics
      [scale=.32,height=.24\textheight,width=.22\textwidth]{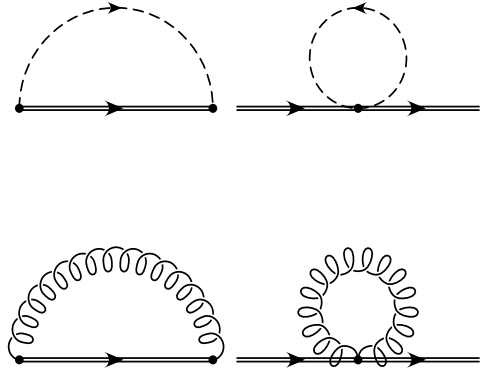}
      \caption{Diagrams contributing to the self energy. The broken lines 
	represent temporal gluons and the curly lines represent 
	spatial gluons.}
      \label{diag}
\vspace{-.4cm}
\end{figure}
\end{center}

The heavy quark's self energy $\Sigma$, represented diagrammatically
in Fig.~\ref{diag}, along with tadpole counterterms from the $u_0$
factors, is defined by writing the inverse quark propagator
$G^{-1}(k)$ in the form
\begin{eqnarray}
  G^{-1}(k) = Q^{-1}(k) - a\Sigma(k)
\end{eqnarray}
\noindent
where $Q^{-1}$ is the free propagator. The self energy can also be
expanded for small $k$ \cite{morn} to give
\begin{eqnarray}
  a\Sigma (k)  \hspace{-.3cm}&\approx&  \hspace{-.3cm}
  \alpha_s \bigg(
  \Omega_0 
  - ik_0 \Omega_1 
  \nonumber \\
  & & \hspace{1cm}+ \,
  \frac{{\bf k}^2}{2\gamma M}\Omega_2 
  + {\bf v.k}\, \Omega_v + \cdots
  \bigg)
  \label{self}
\end{eqnarray}

\vspace{-.5cm}
\noindent
Combining the expansion of $\Sigma$ with the quark propagator, 
to $O(\alpha_s)$, gives
\begin{eqnarray}
  G^{-1}(k) \hspace{-.3cm}&\approx& \hspace{-.3cm}
  - ik_0  -\alpha_s \Omega_0 
  + \alpha_sik_0\Omega_1
  + \frac{{\bf k}^2}{2\gamma M} \nonumber \\
  \,\,&-&\hspace{-.18cm} \alpha_s \frac{{\bf k}^2}{2\gamma M}\Omega_2
  + {\bf v . k} 
  - \alpha_s {\bf v.k}\, \Omega_v
  + \cdots
\end{eqnarray}

\noindent
Re-arranging this expression, we obtain
\begin{eqnarray}
  G^{-1} \hspace{-0.3cm}&=&\hspace{-0.3cm}
  Z_\psi 
  \bigg( -i(k_4 - i\alpha_s\Omega_0) \nonumber\\
  && \hspace{1.5cm} 
  + \frac{{\bf k}^2}{2 \gamma_r M_r } 
  + \frac{{\bf P}_r\cdot{\bf k}}{2 \gamma_r M_r}
  + \cdots \bigg)
\end{eqnarray}

\noindent
The renormalisation parameters are then defined as
\begin{eqnarray}
  Z_\psi \hspace{-.2cm}&=&\hspace{-.2cm}
  1- \alpha_s (\Omega_0 + \Omega_1)\\
  Z_M    \hspace{-.2cm}&=&\hspace{-.2cm}
  1 + \alpha_s (\Omega_2 - \Omega_1) 
    + \alpha_s (\Omega_v - \Omega_1)\,{\bf v}^2 \gamma^2\\
  Z_P    \hspace{-.2cm}&=&\hspace{-.2cm} 
  1 - \alpha_s (\Omega_v - \Omega_2)
\end{eqnarray}
\noindent
where $M_r=Z_M M$ and $P_r=Z_P P$ ($P=\gamma M{\bf v}$). The energy shift is
defined as $E_0 =-\alpha_s \Omega_0$.  We also note that since the
tadpole counterterms are the same for $\Omega_2$ and $\Omega_v$, these
factors will cancel in $Z_P$.  In order to compute the renormalisation
parameters, we need to isolate the coefficients, $\Omega_n$, of
individual terms appearing in Eqn.~\ref{self}.  This is equivalent to
taking derivatives of the diagrams with respect to an appropriate
choice of momentum.

The quark propagator is zero for $t\leq0$, so by choosing the momentum
flow appropriately, poles in the quark propagator can be avoided. The
diagrams have internal loop momentum $q$ and contain a pole in the
gluon propagator in the $q_0$ plane. The integrals are therefore
evaluated analytically in the complex plane ($z=e^{iq_0}$) which leads
to a 3-D integral that can be evaluated numerically using the VEGAS
package. The calculations were performed at $aM=2.0$ with several
different values of the stabilisation parameter and
velocity. These values match those used in the numerical
calculations \cite{foley} which were carried out on Wilson glue configurations 
at $\beta=5.7$. We take $\alpha_s (1/a)=0.35$. The results are 
checked for invariance under  $v_j \rightarrow -v_j$.

\section{PRELIMINARY RESULTS}
\begin{figure}
\vspace{-.3cm}
\hspace{-.65cm}
  \includegraphics
      [scale=.3,height=.325\textheight,width=.49\textwidth]{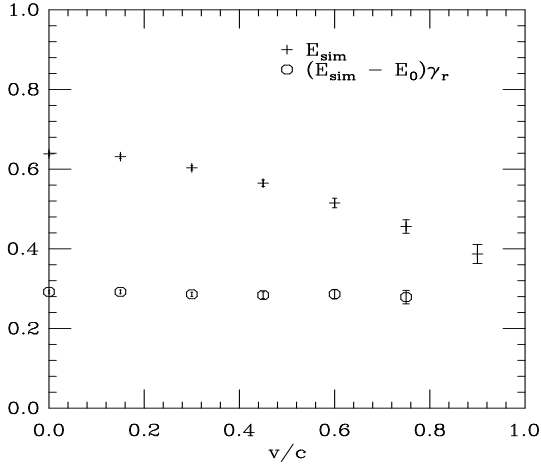}
      \caption{The simulated energy ($E_{\mathrm{sim}}$) and the 
	physical binding energy ($(E_{\mathrm{sim}}-E_0) \gamma_r $)
      for a heavy-light meson with $aM=2$.}
      \label{be}
\vspace{-.9cm}
\end{figure}

Preliminary results for the heavy-light binding energy
$M_{\mathrm{kin}} - Z_M M = (E_{\mathrm{sim}}-E_0) \gamma_r $ 
as a function of the bare velocity are
presented in Fig.~\ref{be}. The simulated energy \cite{foley} clearly
exhibits velocity dependence as $v/c$ approaches 1. After subtraction
of the energy shift, the physical binding energy is independent
of $v$, as it should be.

\begin{figure}
\vspace{-.3cm}
\hspace{-.25cm}
  \includegraphics
      [scale=.3,height=.325\textheight,width=.49\textwidth]
      {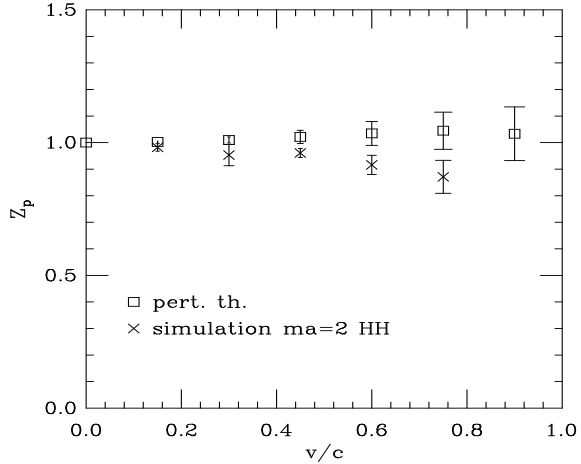}
      \caption{The renormalisation of the external momentum as a 
	function of bare 
	velocity ($v/c$). Simulation errors are statistical only. 
      Perturbative errors are taken as $1 \times \alpha_s^2 v^2$.}
      \label{vel}
\vspace{-.6cm}
\end{figure}

Preliminary values for the renormalisation of the external
momentum $\gamma M {\bf v}$ as a function of the 
bare velocity are presented in Fig.~\ref{vel}. Up to lattice
artefacts, the Lagrangian has an invariance under arbitrary
shifts of momentum between $\gamma M{\bf v}$ and ${\bf k}$, and this protects
$Z_P$ from becoming very different from 1 \cite{pub}. $Z_P$
from perturbation theory and simulations disagree at the level
of systematic errors in this calculation.

\section{CONCLUSIONS}
We have presented preliminary results for the energy shift and
external momentum renormalisation for a heavy quark in moving NRQCD.
Comparison with numerical simulation is encouraging.

The full set of renormalisation parameters will be presented in a
forthcoming publication \cite{pub}. Future work will
incorporate higher order terms in the Lagrangian
and include the renormalisation
of current operators for $B \rightarrow \pi$ decay.

\end{document}